\documentstyle[12pt,epsfig,cite]{article}

\def\be{\begin{equation}}
\def\ee{\end{equation}}
\def\ba{\begin{eqnarray}}
\def\ea{\end{eqnarray}}

\begin{document}

\begin{titlepage}

\renewcommand{\thefootnote}{\fnsymbol{footnote}}


\vspace{0.3cm}

\begin{center}
{\Large\bf Thermodynamic stability of modified Schwarzschild-AdS
black hole in rainbow gravity}
\end{center}

\begin{center}
Yong-Wan Kim\footnote{Electronic address:
ywkim65@gmail.com}$^{1}$, Seung Kook Kim\footnote{Electronic
address: skandjh@seonam.ac.kr}$^{2}$,  and Young-Jai
Park$\footnote{Electronic address: yjpark@sogang.ac.kr}^{3}$\par
\end{center}

\begin{center}

{${}^{1}$Research Institute of Physics and Chemistry,\\ Chonbuk
National University, Jeonju 54896, Korea,}\par {${}^{2}$Department
of Physical Therapy, Seonam University, Namwon 55724, Korea}\par
{${}^{3}$Department of Physics, Sogang University, Seoul 04107,
Korea}\par
\end{center}

\vskip 0.5cm
\begin{center}
{\today}
\end{center}

\vfill

\begin{abstract}
In this paper, we have extended the previous study of the
thermodynamics and phase transition of the Schwarzschild black
hole in the rainbow gravity to the Schwarzschild-AdS black hole
where metric depends on the energy of a probe. Making use of the
Heisenberg uncertainty principle and the modified dispersion
relation, we have obtained the modified local Hawking temperature
and thermodynamic quantities in an isothermal cavity. Moreover, we
carry out the analysis of constant temperature slices of a black
hole. As a result, we have shown that there also exists another
Hawking-Page-like phase transition in which case a locally stable
small black hole tunnels into a globally stable large black hole
as well as the standard Hawking-Page phase transition from a hot
flat space to a black hole.

\end{abstract}

\vskip20pt





\end{titlepage}
\section{ Introduction}
\setcounter{equation}{0}
\renewcommand{\theequation}{\arabic{section}.\arabic{equation}}

The possibility that standard energy-momentum dispersion relations
are modified near the Planck scale is one of the scenarios in
quantum gravity phenomenology
\cite{AmelinoCamelia:1996pj,AmelinoCamelia:2008qg}. Such a
modified dispersion relation (MDR) was also advocated by the study
of nonlinear deformed or doubly special relativity (DSR)
\cite{AmelinoCamelia:2000mn,AmelinoCamelia:2010pd} in which the
Planck length as well as the speed of light is an observer
invariant quantity. In particular, Magueijo and Smolin
\cite{Magueijo:2002am,Magueijo:2002xx} have extended the DSR to
general relativity by proposing that the spacetime background felt
by a test particle would depend on its energy $\omega$. Such an
energy of the test particle deforms the background geometry and
consequently the MDR as
 \begin{equation}\label{MDR2}
 \omega^2 f(\omega/\omega_p)^2-p^2 g(\omega/\omega_p)^2=m^2,
 \end{equation}
where $p$, $m$, $\omega_p$ are the momentum, the mass of the test
particle, and the Planck energy, respectively. Thus, quanta of
different energies see different background geometry, which is
referred to as a rainbow gravity. Since then many efforts have
been devoted to the rainbow gravity related to the gravity and
other stimulated work at the Planck scale
\cite{Liberati:2004ju,Galan:2004st,Galan:2005ju,
Hackett:2005mb,Ling:2006az,
Ling:2006ba,Girelli:2006fw,Ling:2008sy, Garattini:2011hy,
Garattini:2011kp,Garattini:2011fs,Garattini:2012ec,Garattini:2012ca,Majumder:2013mza,
Amelino-Camelia:2013wha,
Awad:2013nxa,Barrow:2013gia,Santos:2015sva,Carvalho:2015omv,Ashour:2016cay}.

In connection with black hole thermodynamics in the rainbow
gravity, there have also been much work with the following rainbow
functions
\cite{AmelinoCamelia:1996pj,AmelinoCamelia:2008qg,Ali:2014xqa,Ali:2014qra,Ali:2014zea,
Gim:2015zra,Hendi:2015hja,Mu:2015qna,Hendi:2015cra,Gim:2015yxa,Gangopadhyay:2016rpl}
\begin{equation}\label{rainbowfunc}
f(\omega/\omega_p)=1, \quad g(\omega/\omega_p)=\sqrt{1-\eta
\left({\omega}/{\omega_p}\right)^n},
\end{equation}
which are one of the most interesting MDRs related to quantum
gravity phenomenology among several other types
\cite{Magueijo:2002am,Garattini:2011hy,Awad:2013nxa,AmelinoCamelia:2005ik,Ling:2005bq,
Ling:2005bp,Galan:2006by,Liu:2007fk,Peng:2007nj,Han:2008sy,Leiva:2008fd,Li:2008gs,
Garattini:2009nq,Liu:2011zd,Liu:2014ema,Garattini:2014rwa,Hendi:2016vux,Hendi:2016aaa,Yadav:2016nfh,Zhao:2016iwi}.
Here, $n$ is a positive integer, $\eta$ a constant of order unity,
and these functions satisfy with $\lim_{\omega \rightarrow 0}
f(\omega/\omega_p) = 1$ and $\lim_{\omega \rightarrow 0}
g(\omega/\omega_p) = 1$ at low energies. In particular, Li {\it et
al.} \cite{Li:2008gs} have obtained the Schwarzschild-AdS black
hole solution in the framework of rainbow gravity with different
rainbow functions from Eq.(\ref{rainbowfunc}), and investigated
thermodynamic stability without the analysis of phase transition.
Recently, Gim and Kim (GK) \cite{Gim:2014ira} have shown that the
Schwarzschild black hole in the rainbow gravity in an isothermal
cavity has an additional Hawking-Page phase transition near the
event horizon apart from the standard one, which is of relevance
to the existence of a locally small black hole.

On the other hand, an approach of the higher dimensional flat
embedding is used to study a local temperature for a freely
falling observer outside black holes
\cite{Brynjolfsson:2008uc,Kim:2013wpa}. And very recently, we have
shown that a local temperature seen by a freely falling observer
depends only on $g(\omega/\omega_p)$ \cite{Kim:2015wwa} so that
the choice of $f(\omega/\omega_p)=1$ makes not only the time-like
Killing vector in the rainbow Schwarzschild black hole as usual,
but also makes the local thermodynamic energy independent of the
test particle's energy.

In this paper, we would extend the GK's work of the Schwarzschild
black hole in the rainbow gravity to the Schwarzschild-AdS
spacetime. In order to study efficiently, we shall describe
thermodynamics by using an event horizon $r_+$ as a variable
instead of the mass $M$ as in the GK's work, since in the
Schwarzschild-AdS black hole it is difficult to solve the event
horizon as a function of the mass. In section 2, black hole
temperature for the Schwarzschild-AdS black hole in the rainbow
gravity will be calculated from the definition of the standard
surface gravity. Then, making use of the MDR and the Heisenberg
uncertainty principle, the energy dependence of a test particle in
black hole temperature will be properly rephrased. And the entropy
will be derived from the first law of thermodynamics. In section
3, we will study local thermodynamic quantities including
temperature, energy and heat capacity in an isothermal cavity with
their various limits for each other's comparison. Furthermore, in
order to clearly reconfirm thermodynamic stability, we also
analyze constant temperature slices of the Schwarzschild(-AdS)
black hole in the rainbow gravity. In section 4, we will study
phase transition between various black hole states and the hot
flat space through investigating free energies of the
Schwarzschild-AdS black hole in the rainbow gravity. Finally,
conclusion and discussion will be given in section 5.

\section{Temperature and entropy of Schwarzschild-AdS black hole in rainbow gravity}
\setcounter{equation}{0}
\renewcommand{\theequation}{\arabic{section}.\arabic{equation}}

Let us consider the modified Schwarzschild-AdS black hole in
rainbow gravity described as \cite{Li:2008gs}
\begin{equation}\label{metric}
ds^2=-\frac{N^2}{f^2(\omega/\omega_p)}dt^2+\frac{1}{g^2(\omega/\omega_p)
N^2}dr^2+\frac{r^2}{g^2(\omega/\omega_p)} d\Omega^2,
\end{equation}
where
\begin{equation}\label{metric1}
N^2 = 1-\frac{2G_0M}{r} + \frac{r^2}{l^2_0}.
\end{equation}
This is a spherically symmetric solution to the modified field
equation in rainbow gravity of
\begin{equation}
 G_{\mu\nu}(\omega/\omega_p)+\Lambda(\omega/\omega_p)g_{\mu\nu}(\omega/\omega_p)
 =8\pi G(\omega/\omega_p) T_{\mu\nu}(\omega/\omega_p)
\end{equation}
in the absence of matter. Here,  $G_{\mu\nu}(T_{\mu\nu})$ is the
Einstein (energy-momentum) tensor,
$G(\omega/\omega_p)(\Lambda(\omega/\omega_p))$ is an
energy-dependent Newton (cosmological) constant, and
$G_0(\Lambda_0=-3/l^2_0)$ is the physical Newton (cosmological)
constant at the low-energy limit of $\omega/\omega_p\rightarrow
0$. Here, the energy $\omega$-dependent constants are related with
the physical ones as
\begin{eqnarray}
G(\omega/\omega_p)&=&\frac{G_0}{g(\omega/\omega_p)},\nonumber\\
\Lambda(\omega/\omega_p)&=&-\frac{3}{l^2(\omega/\omega_p)}=g^2(\omega/\omega_p)\Lambda_0.
\end{eqnarray}
Note that the solution reduces to the usual Schwarzschild-AdS
vacuum solution
\begin{equation}
ds^2=-N^2dt^2+\frac{1}{N^2}dr^2+r^2 d\Omega^2
\end{equation}
in the low energy limit of $\omega/\omega_p\rightarrow 0$. The
mass defined by $N^2=0$ is given by
\begin{equation}
 M(r_+)=\frac{r_+}{2G_0}\left(1+\frac{r^2_+}{l_0^2}\right)
\end{equation}
with the event horizon $r_+$.

Then, the modified Hawking temperature $T_{\rm H}$ is obtained as
\begin{equation}\label{HawkingT}
T_{\rm H}= \frac{\kappa_{\rm H}}{2\pi}
         = \frac{g(\omega/\omega_p)}{f(\omega/\omega_p)}T^0_{\rm H}
\end{equation}
from the surface gravity $\kappa_H$ at the event horizon as
follows
\begin{equation}
\kappa_{\rm H}=-\frac{1}{2}\lim_{r\rightarrow
r_+}\sqrt{\frac{-g^{11}}{g^{00}}}\frac{(g^{11})'}{g^{00}}.
\end{equation}
Here, the standard Hawking temperature $T^0_{\rm H}$ of the
Schwarzschild-AdS black hole is given by
\begin{equation}
T^0_{\rm H} =
\frac{1}{4\pi}\left(\frac{1}{r_+}+\frac{3r_+}{l_0^2}\right).
\end{equation}
Making use of the explicit rainbow functions (\ref{rainbowfunc}),
the black hole temperature can be written as
\begin{equation}
\label{temp} T_{\rm H} =\sqrt{1-\eta(\omega/\omega_p)^n}T^0_{\rm
H}
\end{equation}
so that the temperature depends on the energy $\omega$ of a probe.
\begin{figure*}[t!]
   \centering
   \includegraphics{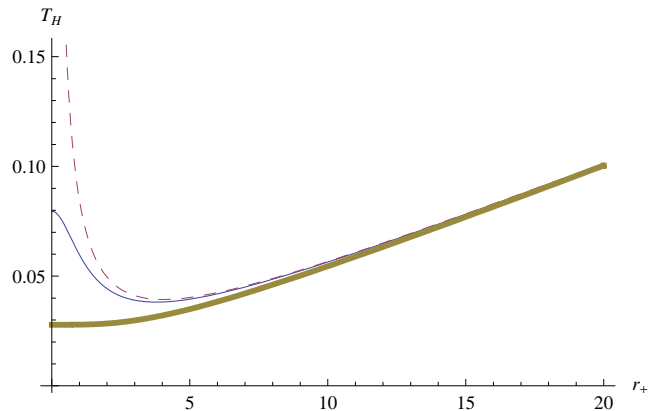}
\caption{The modified Hawking temperatures of the
Schwarzschild-AdS black hole in the rainbow gravity for $\eta=1$
(solid line) and for the upper bound of $\eta=49/6$ (thick line)
with $l_0=7$, $G_0=1$, and the standard Hawking temperature
(dashed line) with $l_0=7$.}
 \label{fig1}
\end{figure*}

Now, in order to eliminate the $\omega$ dependence of the probe in
the modified Hawking temperature (\ref{temp}), one can use the
Heisenberg uncertainty principle (HUP) as in
Ref.~\cite{Gim:2014ira}. In the vicinity of the black hole
surface, an intrinsic position uncertainty $\Delta x$ of the probe
of order of about the event horizon $r_+$ leads to momentum
uncertainty of order of $p$ \cite{Adler:2001vs} as
 \begin{equation}\label{deltap}
p= \Delta p \sim  \frac{1}{r_+}.
 \end{equation}
Plugging the momentum uncertainty into  the MDR (\ref{MDR2}), one
can determine the energy $\omega$. We explicitly show this by
choosing $n=2$ in the rainbow functions (\ref{rainbowfunc})
without loss of generality. Then, the energy for massless particle
can be solved as
\begin{equation}\label{omega2}
\omega =  \frac{\omega_p}{\sqrt{\eta+{r^2_+}\omega_p^2 }}.
\end{equation}
Therefore, one can rewrite the black hole temperature (\ref{temp})
as
\begin{equation}\label{T2}
T_{\rm H}=\frac{r_+}{\sqrt{{r^2_+}+\eta G_0}}T^0_{\rm H},
\end{equation}
with $\omega_p^2=1/G_0$. When $\eta=0$, it is just the standard
Hawking temperature of the Schwarzschild-AdS black hole. Note that
as $r_+\rightarrow 0$, it becomes finite as $T_H=1/(4\pi\sqrt{\eta
G_0})$. This result implies that the standard divergent Hawking
temperature of the Schwarzschild-AdS black hole could be
regularized in the rainbow gravity as like in the Schwarzschild
case \cite{Gim:2014ira}. In Fig. \ref{fig1}, we have plotted the
modified and the standard Hawking temperatures showing that the
former is finite at $r_+=0$ due to the rainbow gravity effect
while the latter blows up.

It seems appropriate to comment that the modified Hawking
temperature has its minimum
\begin{equation}
T^m_{\rm H}=\frac{\sqrt{3}}{2\pi l_0}\sqrt{1-\frac{3\eta
G_0}{l^2_0}}
\end{equation}
at $r_+=\frac{l_0}{\sqrt{3}}\sqrt{1-\frac{6\eta G_0}{l_0^2}}$.
Moreover, the upper bound for the parameter $\eta<l_0^2/6 G_0$ is
required for $r_+$ being real.
In Fig. \ref{fig1}, the thick line is for $\eta=l^2_0/6G_0$ where
$r_+$ is zero and the curve has its minimum.

Next, from the first law of black hole thermodynamics, one can
obtain the entropy as
\begin{equation}\label{S}
S=\int\frac{dM}{T_H}=\frac{\pi r_+}{G_0}\sqrt{r^2_+ + \eta G_0} +
\pi \eta \sinh^{-1}\left(\frac{r_+}{\eta G_0}\right).
\end{equation}
Here, we have chosen the integration constant to be satisfied with
$S\rightarrow 0$ as $r_+\rightarrow 0$. This is the exactly same
form with the entropy of the Schwarzschild black hole in the
rainbow gravity~\cite{Gim:2014ira}, but explicitly different in
$r_+$. This is nothing new since we know that the standard
Schwarzschild black hole and the Schwarzschild-AdS black hole also
have the same form of entropy of $S=\pi r^2_+$, but different in
$r_+$. Moreover, in the standard Schwarzschild-AdS limit of
$\eta=0$, the entropy (\ref{S}) becomes one-fourth of the area of
the event horizon $S=A/4G_0$. Note that the next leading order is
logarithmic as $S \approx A/4G_0+\frac{1}{2}\pi\eta\ln(A/4)$,
which is reminiscent of quantum correction to the entropy
\cite{Fursaev:1994te,Kaul:2000kf,Das:2001ic,Chatterjee:2003uv,Wang:2008zzc}.
From these, one can see that the rainbow metric contributes to the
quantum corrected metric.

At this stage, it is appropriate to comment on the choice of $n=2$
in the rainbow functions (\ref{rainbowfunc}), which makes us to
analytically solve the MDR (\ref{MDR2}) and eventually gives the
logarithmic correction to the entropy. If not, it would be
difficult to solve the MDR first, and then have another forms of
correction without the logarithmic term to the entropy
\cite{Mu:2015qna}.

\section{Schwarzschild-AdS black hole thermodynamics in rainbow gravity}
\setcounter{equation}{0}
\renewcommand{\theequation}{\arabic{section}.\arabic{equation}}

Now, let us study Schwarzschild-AdS black hole thermodynamics in
rainbow gravity. In order for the study to be focused and feasible,
we take the rainbow functions (\ref{rainbowfunc}) with $n=2$.

We would like to consider a local observer who is at rest at the
radius $r$, which we mean to introduce a spherical cavity
enclosing the Schwarzschild-AdS black hole. Then, the local
temperature $T_{\rm{loc}}$ seen by the local observer can be
obtained as
\begin{equation}\label{locT}
 T_{\rm{loc}} =\frac{T_H}{\sqrt{-g_{00}}}
              = \frac{\frac{1}{4\pi}\left(\frac{1}{r_+}+\frac{3r_+}{l_0^2}\right)\sqrt{\frac{r^2_+}{r^2_+ + \eta G_0}}}
                     {\sqrt{1-\frac{r_+}{r}\left(1+ \frac{r^2_+}{l^2_0}\right)+\frac{r^2}{l_0^2}}}
\end{equation}
by implementing the redshift factor of the
metric~\cite{Tolman:1930zza}.

Before proceeding further, let us briefly look into the limits of
the modified Hawking temperature (\ref{T2}) and local temperature
(\ref{locT}) of the Schwarzschild-AdS black hole in the rainbow
gravity enclosed in a cavity. First, by turning off the rainbow
gravity, in the Schwarzschild-AdS limit of $\eta\rightarrow 0$,
the standard Hawking and local temperatures will be
\begin{eqnarray}
\label{HTAdS} T^{\rm SAdS}_{\rm H} &=&
 \frac{1}{4\pi}\left(\frac{1}{r_+}+\frac{3r_+}{l_0^2}\right),\\
\label{LTAdS} T^{\rm SAdS}_{\rm loc} &=&
 \frac{\frac{1}{4\pi}\left(\frac{1}{r_+}+\frac{3r_+}{l_0^2}\right)}
                     {\sqrt{1-\frac{r_+}{r}\left(1+
                     \frac{r^2_+}{l^2_0}\right)+\frac{r^2}{l_0^2}}},
\end{eqnarray}
and in the Schwarzschild limit of $l^2_0\rightarrow\infty$ as well
as $\eta\rightarrow 0$, those become
\begin{eqnarray}
 \label{HT} T^{\rm Sch}_{\rm H} &=& \frac{1}{4\pi r_+},\\
 \label{LT} T^{\rm Sch}_{\rm loc} &=&  \frac{1}{4\pi r_+\sqrt{1-\frac{r_+}{r}}}.
\end{eqnarray}
\begin{figure*}[t!]
   \centering
   \includegraphics{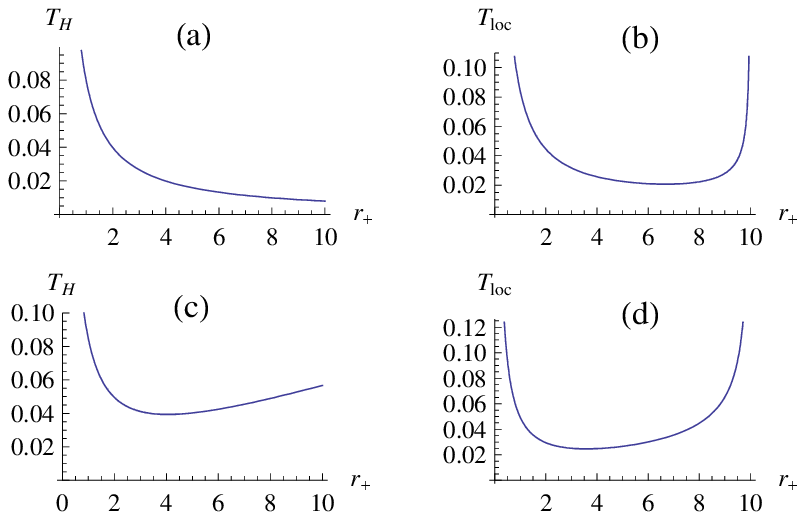}
\caption{Temperatures for (a) the Schwarzschild black hole, (b)
the Schwarzschild black hole enclosed in a cavity, (c) the
Schwarzschild-AdS black hole, and (d) the Schwarzschild-AdS black
hole enclosed in a cavity.}
 \label{fig2}
\end{figure*}
\begin{figure*}[t!]
   \centering
   \includegraphics{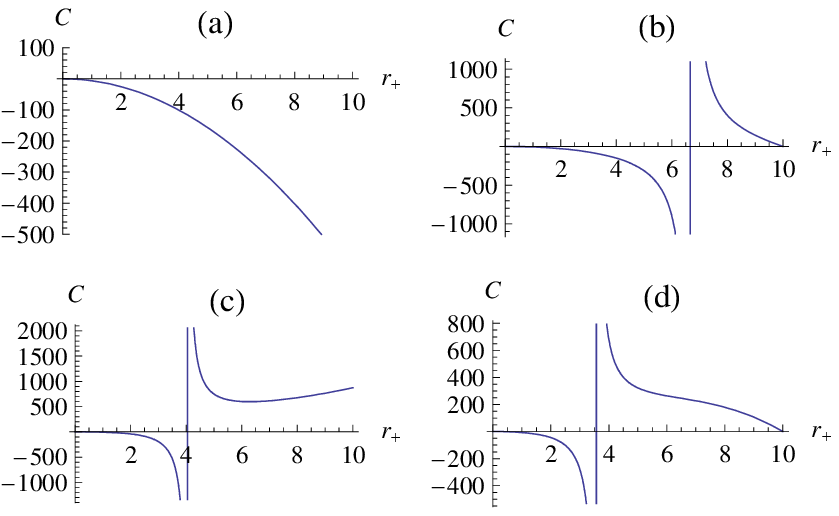}
\caption{Specific heats for (a) the Schwarzschild black hole, (b)
the Schwarzschild black hole enclosed in a cavity, (c) the
Schwarzschild-AdS black hole, and (d) the Schwarzschild-AdS black
hole enclosed in a cavity.}
 \label{fig3}
\end{figure*}
\begin{figure*}[t!]
   \centering
   \includegraphics{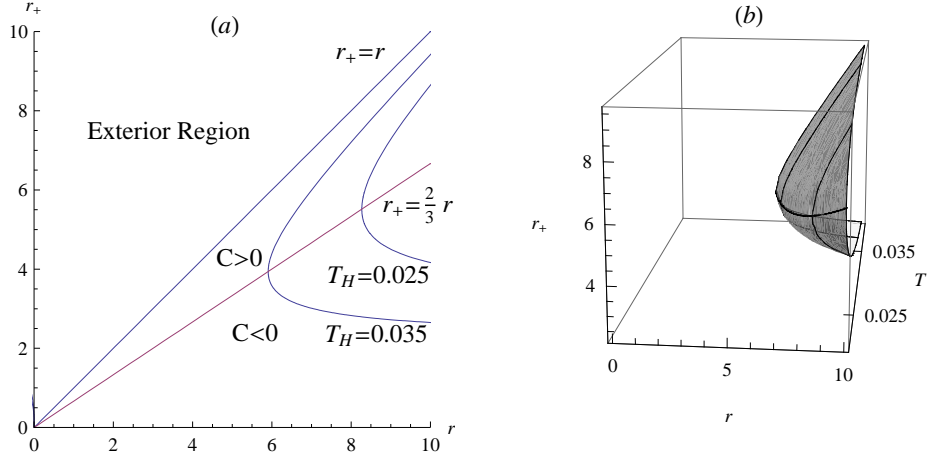}
\caption{Constant temperature slices of the Schwarzschild black
hole confined within a cavity.}
 \label{fig4}
\end{figure*}

For the sake of detailed discussion, their corresponding specific
heats  can also be written down as
\begin{eqnarray}
\label{spads}C^{\rm SAdS}&=& -\frac{2\pi r^2_+}{G_0}\left(\frac{l^2_0+3r^2_+}{l^2_0-3r^2_+}\right),\\
\label{Lspads}C^{\rm SAdS}_{\rm loc}&=&\frac{4\pi
r^2_+(r-r_+)(l^2_0+r^2+rr_++r^2_+)(l^2_0+3r^2_+)}{G_0[r_+(3l^4_0+2l^2_0r^2_++3r^4_+)-2r(l^2_0+r^2)(l^2_0-3r^2_+)]}
\end{eqnarray}
for the Schwarzschild-AdS black hole, and
\begin{eqnarray}
\label{spsch}C^{\rm Sch}&=& -\frac{2\pi r^2_+}{G_0},\\
\label{Lspsch}C^{\rm Sch}_{\rm loc}&=& -\frac{4\pi
r^2_+}{G_0}\left(\frac{r-r_+}{2r-3r_+}\right)
\end{eqnarray}
for the Schwarzschild black hole. Here, the subscript `loc'
indicates that specific heat is obtained from a black hole
enclosed in a cavity.

Figure \ref{fig2}(a) shows the standard Hawking temperature
(\ref{HT}) of the Schwarzschild black hole which becomes zero as
$r_+$ goes to infinity, while it diverges as $r_+$ goes to zero.
If it is enclosed in a cavity, the local temperature at the cavity
(\ref{LT}) goes up to infinity. This cavity makes the black hole
thermodynamically well defined. Then, this thermodynamic stability
can be easily seen if one examines the specific heat of $T(dS/dT)$
as in Fig. \ref{fig3}. In Fig. \ref{fig3}(a), one sees the
negative specific heat as given in Eq.~(\ref{spsch}), which
indicates that the Schwarzschild black hole is unstable. On the
other hand, as seen in Fig. \ref{fig3}(b), the Schwarzschild black
hole in the cavity has an asymptote at $r_+=2r/3$, which is
obtained from the condition of $dT_{\rm loc}/dr_+=0$, and if $r_+$
is larger than $2r/3$, the black hole is stable. More precisely,
if $r_+ T_{\rm loc}\ge \sqrt{3}/4\pi$, there are two possible
solutions in which the larger one is stable while the smaller one
is unstable as shown in Fig. \ref{fig2}(b) and Fig. \ref{fig3}(b).
The equality takes place if the temperature is the minimum at
$r_+=2r/3$. Below the minimum temperature there is no black hole
solution. These can be reconfirmed by analyzing constant
temperature slices \cite{Prestidge:1999uq} of the Schwarzschild
black hole confined in a cavity in Figs. \ref{fig4}(a), (b), which
are obtained by solving Eq. (\ref{LT}) in term of $r_+$ at a
constant temperature. In the diagram, one can easily see that
thermodynamic stability is in between $2r/3<r_+<r$. The turning
points of the curves satisfying with $r_+ T_{\rm
loc}=\sqrt{3}/4\pi$ remain on the line $r_+=2r/3$. Two solutions
of the Schwarzschild black hole in the cavity are possible to the
right of this point, one above the line and one below it. They
would meet if a given temperature equals to the minimum
temperature. For higher cavity temperature, the constant
temperature curve shifts to the left, and for lower cavity
temperature it shifts to the right.

Compared with the standard Schwarzschild black hole which is
asymptotically flat, the temperatures and specific heats of the
Schwarzschild-AdS black hole, which is not asymptotic flat, are
depicted in Figs. \ref{fig2}-\ref{fig3} (c) and (d). When compared
the Hawking temperature in Fig. \ref{fig2}(c) with the local
temperature in Fig. \ref{fig2}(b), one can see that they show the
same behavior with each other except that in the case of Fig.
\ref{fig2}(c) the cavity seems to be located at infinity. Thus,
the asymptotically AdS geometry naturally plays the role of a
cavity. The real positive solutions of the Schwarzschild-AdS black
hole can be obtained when $r_+T_{\rm H}\ge 1/2\pi$, where the
equality takes place if the temperature is the minimum at
$r_+=l_0/\sqrt{3}$ ($\eta=0$ case). The larger solution is locally
stable and has a positive specific heat, while the smaller
solution is unstable and has a negative specific heat as in Fig.
\ref{fig2}-\ref{fig3} (c).

If the Schwarzschild-AdS black hole is enclosed in a cavity, the
local temperature (\ref{LTAdS}) and specific heat (\ref{Lspads})
are depicted as in Figs. \ref{fig2}-\ref{fig3} (d). As $r_+$
approaches the radius $r$ of the cavity, the local temperature
goes up to infinity, and the specific heat to zero. This is also
the case for the Schwarzschild black hole enclosed in the cavity,
which can be seen through the factor of $C\sim (r-r_+)$ in Eqs.
(\ref{Lspads}) and (\ref{Lspsch}), while the specific heat of the
standard Schwarzschild-AdS black hole diverges as in Fig.
\ref{fig3} (c).

\begin{figure*}[t!]
   \centering
   \includegraphics{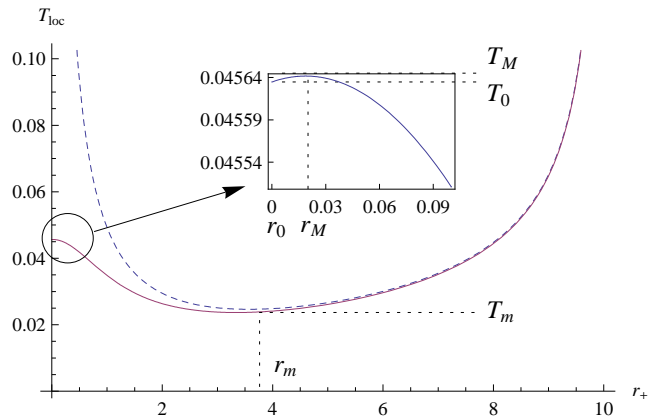}
\caption{Local temperature of the Schwarzschild-AdS black hole in
the $\eta=1$ rainbow gravity enclosed in a cavity (solid line) and
the standard local temperature of the Schwarzschild-AdS black hole
(dotted line) with $r=10$, $l=7$. Here, $T_m$ is globally minimum,
$T_M$ is locally maximum, and $T_0$ is the temperature at
$r_+=0$.}
 \label{fig5}
\end{figure*}

Now, let us go back to the local temperature (\ref{locT}), which
is seen by the observer at $r$, is depicted in Fig. \ref{fig5}
where the local temperature of the Schwarzschild-AdS black hole in
the rainbow gravity enclosed in a cavity was also drawn for the
purpose of comparison, which is divergent as $r_+\rightarrow 0$.
The Fig. \ref{fig5} is also different from the Fig. \ref{fig1} in
that this is the local temperature seen by a local observer at $r$
but not the modified Hawking temperature seen by an observer at
the asymptotic infinity. One can see in the figure that for a
fixed $r$ the local temperature is divergent as $r_+$ approaches
$r$, which is the same with the local temperature of the
Schwarzschild-AdS black hole. We also observe that there exist a
global minimum temperature $T_m$ at $r=r_m$, and a local maximum
temperature $T_M$ at $r=r_M$. Moreover, the temperature remains
finite as $r_+\rightarrow 0$ with $T_0=1/(4\pi\sqrt{\eta
G_0(1+r^2/l^2)})$. Thus, as a result of the introduction of the
rainbow gravity, one can observe both the existence of the local
maximum temperature near the origin and the finiteness of the
local temperature at the origin, which were first shown in
Ref.~\cite{Gim:2014ira}. However, we also note that these were
originally absent in the modified Hawking temperature (\ref{T2})
as in Fig. \ref{fig1}.

Next, making use of the entropy (\ref{S}) and the local
temperature (\ref{locT}), the thermodynamic first law yields the
energy $E_{\rm{tot}}$ as
\begin{eqnarray}\label{Etot}
E_{\rm{tot}} &=& \int^{r_+}_{0} T_{\rm{loc}} dS  \nonumber \\
             &=& \frac{r}{G_0}\left(\sqrt{1+\frac{r^2}{l_0^2}}
                 -\sqrt{\left(1-\frac{r_+}{r}\right)+\frac{1}{ l_0^2}
                 \left(r^2-\frac{r^3_+}{r}\right)}\right).
\end{eqnarray}
\begin{figure*}[t!]
   \centering
   \includegraphics{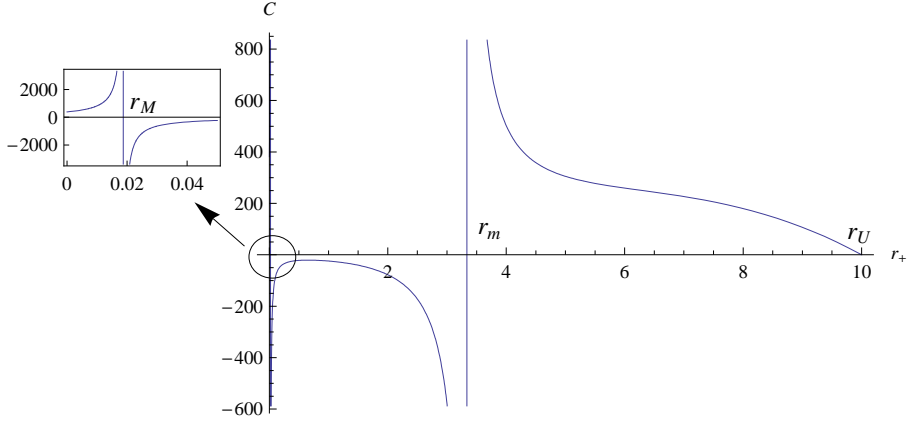}
\caption{Heat capacity of the Schwarzschild-AdS black hole in the
$\eta=1$ rainbow gravity enclosed in the cavity with $r=10$,
$l=7$, $G_0=1$,  where $r_U$ is the upper bound of the event
horizon $r_+$.}
 \label{fig6}
\end{figure*}
To investigate thermodynamic stability of the Schwarzschild-AdS
black hole in the rainbow gravity, we calculate the heat capacity
as
\begin{equation}\label{capacity}
 C =\frac{\partial E_{\rm{tot}}}{\partial T_{\rm{loc}}}
   = \frac{4\pi r^2_+(r-r_+)(l_0^2+r^2+rr_++r^2_+)(l_0^2+3r^2_+)(1+\eta\frac{G_0}{r^2_+})^\frac{3}{2}}
          {G_0H(r_+,r,\eta,G_0)}
\end{equation}
with
\begin{eqnarray}
 H(r_+,r,\eta,G_0)&=& r_+(3l^4_0+2l^2_0r^2_++3r^4_+)-2r(l^2_0+r^2)(l^2_0-3r^2_+) \nonumber\\
                   &-& \eta G_0\left[3(r^3_+- 4
                   r^3)+6l_0^2(r_+-2r)-\frac{l_0^4}{r_+}\right].
\end{eqnarray}
When $\eta\rightarrow 0$, it becomes the specific heat
(\ref{Lspads}) for the standard Schwarzschild-AdS black hole in
the cavity, and when taking the radius $r$ of the cavity to
infinity, it becomes the specific heat (\ref{spads}) for the
Schwarzschild-AdS black hole. Moreover, when
$l_0\rightarrow\infty$ as well as $\eta\rightarrow 0$, it reduces
to the specific heat (\ref{Lspsch}) for the Schwarzschild black
hole in the cavity, and with the radius of the cavity as
$r\rightarrow\infty$, it becomes the specific heat (\ref{spsch})
for the standard Schwarzschild black hole. In Fig. \ref{fig6} the
specific heat is plotted as a function of $r_+$ and it shows three
qualitatively different regions, two stable and one unstable
states of the black hole. Specifically, when $r_+>r_m$, the
specific heat is positive so that the black hole is stable, which
we will call the large stable black hole (LSB), when
$r_M<r_+<r_m$, it is negative, so the black hole unstable, the
intermediate unstable black hole (IUB), and when $0<r_+<r_M$, it
is again positive, and we call it as the small stable black hole
(SSB). Note that the SSB appears in the very vicinity of the
vanishing event horizon.

\begin{figure*}[t!]
   \centering
   \includegraphics{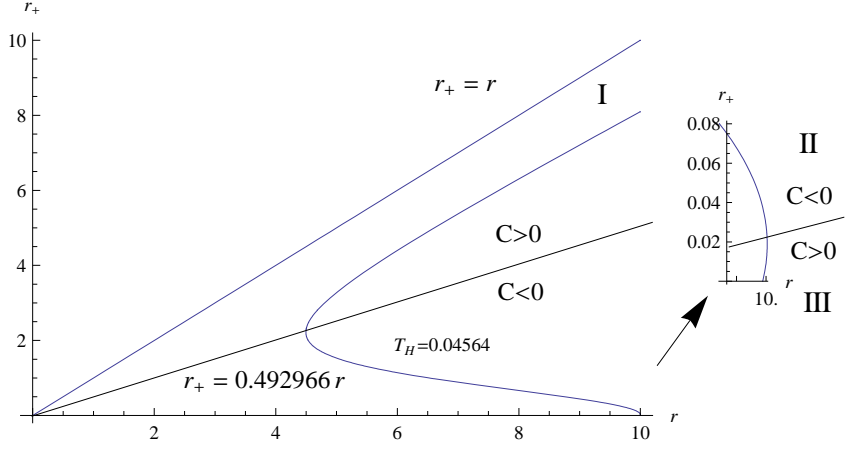}
\caption{Constant temperature slice of the Schwarzschild-AdS black
hole in the rainbow gravity confined within a cavity. Here, we
choose $T=0.04564$ as an external cavity temperature where there
exist three black hole states. }
 \label{fig7}
\end{figure*}

In Fig. \ref{fig7}, it shows constant temperature slices of the
Schwarzschild-AdS black hole in the rainbow gravity confined in
the cavity. Compared with Fig. \ref{fig4}, one can reconfirm that
there is also a fine structure near the vanishing event horizon
due to the appearance of the local maximum temperature as well as
the finiteness of the temperature at the origin, which is enlarged
separately in the right-hand side of the diagram. From this
figure, we see that the thermodynamic states of the
Schwarzschild-AdS black hole in the rainbow gravity confined in
the cavity are divided into three regions; region I with positive
specific heat which corresponds to the LSB, region II with
negative specific heat to the IUB, and region III again with
positive specific heat to the SSB, respectively.

\section{Free energy and phase transition}
\setcounter{equation}{0}
\renewcommand{\theequation}{\arabic{section}.\arabic{equation}}

Now, let us study thermodynamic phase transition
\cite{Hawking:1982dh,York:1986it, Son:2012vj}. The on-shell free
energy of the Schwarzschild-AdS black hole in the rainbow gravity
enclosed in a cavity is obtained  by the use of the local
temperature $T_{\rm loc}$ in Eq.(\ref{locT}) and the thermodynamic
energy $E_{\rm tot}$ in Eq.(\ref{Etot}), explicitly as
\begin{eqnarray}\label{Fon}
 F_{\rm{on}}&=& E_{\rm tot}-T_{\rm loc}S \nonumber\\
            &=& \frac{r}{G_0}\left(\sqrt{1+\frac{r^2}{l^2_0}}
               -\sqrt{\left(1-\frac{r_+}{r}\right)+\frac{1}{
               l_0^2}\left(r^2-\frac{r^3_+}{r}\right)}\right)\nonumber\\
            &-&\frac{r_+\left(1+\frac{3r^2_+}{l^2_0}\right)}
               {4G_0\sqrt{\left(1-\frac{r_+}{r}\right)
                 +\frac{1}{l_0^2}\left(r^2-\frac{r^3_+}{r}\right)}}
                  \left(1+\frac{\eta G_0\sinh^{-1}\left(\frac{r_+}{\eta G_0}\right)}
                  {r^2_+\sqrt{1+\frac{\eta G_0}{r^2_+}}}\right),\nonumber\\
\end{eqnarray}
where $S$ is the entropy in Eq.(\ref{S}). When $\eta =0$, it
becomes the on-shell free energy of the Schwarzschild-AdS black
hole enclosed in a cavity. Furthermore, when $l_0\rightarrow
\infty$, it recovers the on-shell free energy of the Schwarzschild
black hole enclosed in a cavity. In order to understand phase
transition, we also need to introduce off-shell free energy, which
is composed of a set of saddle points of the on-shell free
energies as
\begin{equation}
 F_{\rm{off}}= E_{\rm tot}-T S.
\end{equation}
Here, $E_{\rm tot}$ and $S$ are the same as before, while $T$ is
an external temperature of heat reservoir to control phase
transition.
\begin{figure*}[t!]
   \centering
   \includegraphics{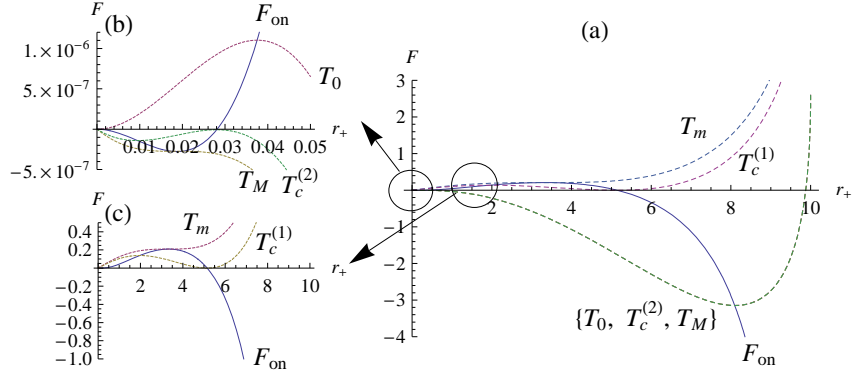}
\caption{On-shell (solid line) and off-shell (dashed-line) free
energies $F(r_+)$ of the Schwarzschild-AdS black hole in the
$\eta=1$ rainbow gravity enclosed in the cavity with $r=10$,
$l=7$, $G=1$, where $T_m$ is global minimum, $T_M$ local maximum
temperature, $T_0$ at $r_+=0$, and $T^{(1)}_c$, $T^{(2)}_c$ phase
transition temperatures.}
 \label{fig8}
\end{figure*}
\begin{figure*}[t!]
   \centering
   \includegraphics{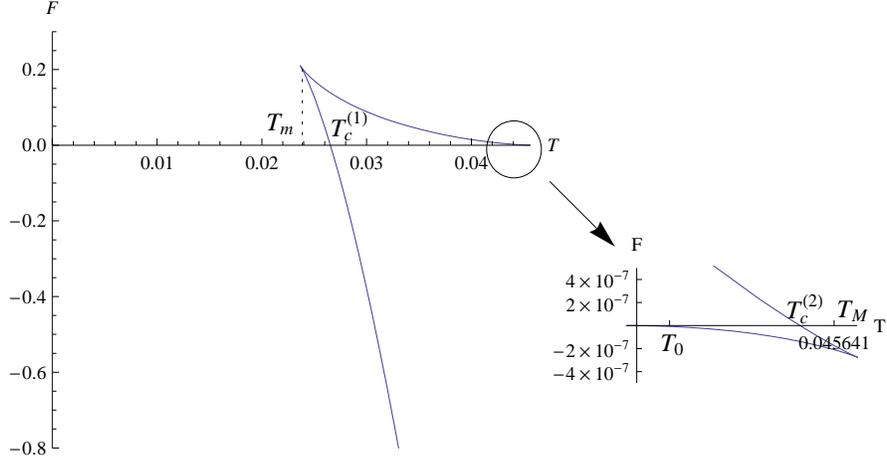}
\caption{On-shell free energy $F(T)$ in terms of $T$ with $r=10$,
$l=7$, $G_0=1$, $\eta=1$. The circle is magnified to show fine
detail which structure comes from the rainbow gravity. }
 \label{fig9}
\end{figure*}

In Fig. \ref{fig8}, we have plotted both the on-shell
$F_{on}(r_+)$ and off-shell $F_{off}(r_+)$ free energies as a
function of $r_+$. This picture is helpful when one considers
thermodynamic stability. On the other hand, in Fig. \ref{fig9}, we
have drawn the on-shell free energy as a function of the
temperature $T$, which helps us to understand phase as temperature
changes. Some regions including near the vanishing event horizon
are magnified to show fine details in Figs. \ref{fig8} and
\ref{fig9}.

Firstly, the situation is mainly divided by two as shown in Fig.
\ref{fig8}(b) and Fig. \ref{fig8}(c), respectively. First, in Fig.
\ref{fig8} (c), which describes the well-known Hawking-Page phase
transition, when $T<T_m$, there are no black holes but a pure
thermal radiation phase which has zero free energy. When $T=T_m$,
the free energy exhibits an inflexion point in Fig. \ref{fig8}(c)
at $r_+=r_m$ where the specific heat is ill-defined, but a single
unstable black hole is formed which eventually decays into a pure
thermal radiation. As the temperature goes up in between
$T_m<T<T^{(1)}_c$, there are two black holes that can be in
equilibrium with a thermal radiation, the small black hole is
formed at a local maximum of the off-shell free energy, while the
large black hole is formed at a local minimum. However, the small
black hole, which corresponds to the IUB with the negative
specific heat, is locally unstable and so decays either into a
thermal radiation or to the large black hole. On the other hand,
the large black hole corresponding to the LSB with positive free
energy is not globally stable, which is locally stable in between
$T_m<T<T^{(1)}_c$ though, so by the black hole evaporation it
would reduce its free energy. In brief, the two small and large
black hole states in this temperature range are less probable than
a pure thermal radiation. When the temperature becomes
$T^{(1)}_c$, the large black hole is at local minimum with zero
free energy, while the small black hole keeps in locally unstable.
When the temperature is in the region of $T^{(1)}_c<T<T_0$, there
are still two black holes that the large black hole is now in
globally stable, which has both positive heat capacity and
negative free energy, but the small black hole has negative heat
capacity and positive free energy so that it is unstable to decay
the globally stable large black hole state. In short, below the
phase transition temperature of $T^{(1)}_c$, it is more probable a
thermal radiation while above $T^{(1)}_c$ more probable a large
black hole so that there exists the Hawking-Page phase transition
between them.

Secondly, in Fig. \ref{fig8} (b), which shows off-shell free
energies near the vanishing event horizon, one observes new mixed
phase transition of the Schwarzschild-AdS black hole in the
rainbow gravity enclosed in a cavity. Note that compared with the
previous case, the phenomenon occurs in relatively extremely tiny
range of temperature between $T_0<T^{(2)}_c<T_M$ where three black
hole states exist. One of them is the SSB and the others are the
same as before, the IUB and LSB. When $T_0<T<T^{(2)}_c$, the IUB
which free energy is at local maximum with positive value and
negative specific heat, is unstable so decays either into a
thermal radiation or to the SSB/LSB. When the temperature is
$T^{(2)}_c$, the IUB has zero free energy. When the temperature is
in between $T^{(2)}_c<T<T_M$, all the three black holes are in
stable states. When $T=T_M$, two stable black holes remains.
However, regardless how stable the SSB is, it is the LSB with much
lower free energy so that the SSB eventually decays into the LSB.
Finally, when $T>T_M$, only the LSB exists.

\section{Discussion}
\setcounter{equation}{0}
\renewcommand{\theequation}{\arabic{section}.\arabic{equation}}

In this paper, we have studied local thermodynamics including its
phase transition of the Schwarzschild-AdS black hole in the
rainbow gravity enclosed in a cavity subject to the MDR. The black
hole temperature in the rainbow gravity depends on the energy
$\omega$ of a probe provided by the rainbow functions
(\ref{rainbowfunc}), and making use of the HUP and deploying the
MDR, we have derived the modified Hawking temperature which is
finite at $r_+=0$. It implies that the divergent standard Hawking
temperature of the Schwarzschild-AdS black hole is regularized in
the rainbow gravity. Moreover, the parameter $\eta$ is found to
have the upper limit given by the Newton and the cosmological
constants where $r_+$ is zero and the modified Hawking temperature
has its minimum value, in contrast to the case of the
Schwarzschild black hole in the rainbow gravity having no upper
bound \cite{Gim:2014ira}.

We have summarized in Fig. \ref{fig2} the temperatures and
specific heats of the Schwarzschild(-AdS) black hole with/without
a cavity which are obtained from taking limits of the
Schwarzschild-AdS black hole in the rainbow gravity enclosed in a
cavity. As a result, from Figs. \ref{fig2}(b), (c) and Figs.
\ref{fig3}(b), (c), we have observed again the well-known facts
that the Schwarzschild-AdS black hole has the similar
thermodynamic behavior as the Schwarzschild black hole in a
cavity; for a given temperature of $T>T_m$, there are two black
holes, the larger one is stable and the small one is unstable,
needless to say, including the Hawking-Page phase transition from
hot flat space to a black hole at the critical temperature,
$T^{(1)}_c$ for our case.

However, for the Schwarzschild-AdS black hole in the rainbow
gravity, due to deformation of temperature near the vanishing
event horizon, the modified local Hawking temperature is quite
different from the standard Hawking temperature of the
Schwarzschild-AdS black hole in a cavity as shown in Fig.
\ref{fig5} where the former is finite even at $r=r_+$ while the
latter is divergent. As a result, it is shown that there exists an
additional stable tiny black hole together with the small black
hole above $T_0$. Moreover, it is also shown that there exists an
additional critical temperature $T^{(2)}_c$ at which the locally
stable tiny black hole tunnels into the stable large black hole
with the finite transition probability seen from Figs. \ref{fig8}
and \ref{fig9} of the on-shell and off-shell free energies.

Finally, it is appropriate to comment that in the presence of the
cosmological constant the HUP in Eq. (\ref{deltap}), which is used
to eliminate the energy dependence of the probe in the modified
Hawking temperature, is corrected by the extended uncertainty
relation~\cite{Bolen:2004sq,Park:2007az}. Therefore, as a further
investigation, it will be interesting to analyze the thermodynamic
quantities and phase transition in the rainbow gravity by using
the generalized uncertainty principle
\cite{Kempf:1994su,Garay:1994en,Li:2002xb,Myung:2006qr,Kim:2007if}.


\section*{Acknowledgments}
We would like to thank Prof. W. Kim and Mr. Y. Gim for
helpful discussions.


\end{document}